\documentclass{article}
 \usepackage{amssymb,amsmath,texdraw}

\begin{document}

 \def\epsilon{\varepsilon}
 \def\phi{\varphi}
 \def\kappa{\varkappa}
 \def\Re{{\rm Re}}

\def\R{{\mathbb R}}
 \def\C{{\mathbb C}}
  \def\ge{\geqslant}
  \def\le{\leqslant}
  \def\Z{{\mathbb Z}}

  \def\frak{\mathfrak}

  \def\fP{{\frak P}}
\def\fQ{{\frak Q}}
\def\fR{{\frak R}}
\def\fU{{\frak U}}
 \def\fp{{\frak p}}
\def\fq{{\frak q}}
\def\fr{{\frak r}}
\def\fL{{\frak L}}
\def\fu{{\frak u}}
\def\fv{{\frak v}}
\def\fw{{\frak w}}

\def\tr{\triangledown}
\def\chtr{\blacktriangledown}

\def\cM{{\cal M}}
\def\cU{{\cal U}}
\def\cV{{\cal V}}
\def\cL{{\cal L}}
\def\cW{{\cal W}}
\def\cS{{\cal S}}

 \def\bm{{\mathbf m}}

   \def\Pol{{\rm Pol}_{\R^*}}
   \def\Poll{{\rm Pol}}
  \def\graph{{\rm graph}}
   \def\Gms{{\rm Gms}}
   \def\fin{{\rm fin}}
\def\Ams{{\rm Ams}}

   \def\Dom{{\rm Dom}}
   \def\Im{{\rm Im}}
   \def\PB{{\rm PB}}
   \def\star{{\curlyvee}}

\def\pic{
  \begin{picture}(300,100)
\put(0,10){
\put(-20,38){$M:$}
\put(20,70){$\phi$}
\thicklines
\put(0,40){\line(1,0){280}}
\thinlines
\put(15,40){\circle{4}}
\put(15,80){\vector(0,-1){37}}
\put(15,55){\circle*{2}}
\put(15,60){\circle*{2}}
\put(15,65){\circle*{2}}
\put(15,70){\circle*{2}}
\put(10,30){$m_2$}
\put(0,15){$p_2=4$}
\put(65,40){\circle{4}}
\put(65,80){\vector(0,-1){37}}
\put(65,55){\circle*{2}}
\put(65,60){\circle*{2}}
\put(60,30){$m_1$}
\put(55,15){$p_1=2$}
\put(135,40){\circle{4}}
\put(135,80){\vector(0,-1){37}}
\put(135,55){\circle*{2}}
\put(130,30){$m_3$}
\put(125,15){$p_3=1$}
\put(235,40){\circle{4}}
\put(235,80){\vector(0,-1){37}}
\put(235,55){\circle*{2}}
\put(230,30){$m_7$}
\put(225,15){$p_7=1$}
\put(285,40){$\dots$}
\put(310,40){\circle*{6}}
\put(305,25){$\xi_\infty$}
\put(310,80){\vector(0,-1){35}}
\put(310,55){\circle*{2}}
\put(310,57){\circle*{2}}
\put(310,59){\circle*{2}}
\put(310,61){\circle*{2}}
\put(310,63){\circle*{2}}
\put(310,65){\circle*{2}}
\put(310,67){\circle*{2}}
\put(310,69){\circle*{2}}
\put(310,71){\circle*{2}}
\put(310,73){\circle*{2}}
\put(310,75){\circle*{2}}
\put(310,77){\circle*{2}}
\put(310,79){\circle*{2}}
\put(310,81){\circle*{2}}
}
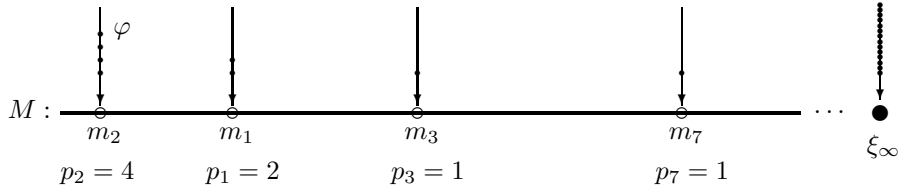
\put(0,8){\sf Picture 1. A configuration.}
\put(0,0){\sf Black points are elements of $Z$, the map
$\phi$ is the \linebreak projection down.}
\end{picture}
}

\def\picc{
\begin{picture}(300,210)
\put(0,10){
\linethickness{1mm}
\put(0,180){\line(1,0){300}}
\put(-20,178){$M:$}
\put(0,150){$\phi$}
\put(0,50){$\psi$}
\put(0,20){\line(1,0){300}}
\put(-20,18){$N:$}
\linethickness{0.3mm}
\multiput(15,135)(0,10){3}
{\multiput(0,0)(70,0){4}{\circle*{3}}}
\multiput(15,35)(0,10){3}
{\multiput(0,0)(70,0){4}{\circle*{3}}}
\linethickness{0.1mm}
\put(15,145){\oval(20,20)[rt]}
\put(15,45){\oval(20,20)[rb]}
\put(25,45){\line(0,1){100}}
\put(85,35){\line(1,1){130}}
\put(215,165){\line(1,0){10}}
\put(225,165){\circle*{3}}
\put(225,145){\oval(20,20)[rt]}
\put(225,45){\oval(20,20)[rb]}
\put(235,45){\line(0,1){100}}
\put(85,45){\line(1,1){40}}
\put(125,85){\line(0,1){40}}
\put(125,125){\line(1,1){30}}
\put(155,135){\oval(20,20)[rt]}
\put(155,45){\oval(20,20)[rb]}
\put(165,45){\line(0,1){90}}
\put(225,135){\oval(20,20)[lt]}
\put(225,55){\oval(20,20)[lb]}
\put(215,55){\line(0,1){80}}
\put(15,45){\line(1,1){40}}
\put(55,85){\line(0,1){40}}
\put(55,125){\line(1,1){30}}
\linethickness{0.3mm}
\put(15,130){\vector(0,1){45}}
\put(15,180){\circle{5}}
\put(15,20){\circle{5}}
\put(15,70){\vector(0,-1){45}}
\put(10,5){$n_1$}
\put(10,190){$m_1$}
\put(85,130){\vector(0,1){45}}
\put(85,180){\circle{5}}
\put(85,20){\circle{5}}
\put(85,70){\vector(0,-1){45}}
\put(80,5){$n_2$}
\put(80,190){$m_2$}
\put(155,130){\vector(0,1){45}}
\put(155,180){\circle{5}}
\put(155,20){\circle{5}}
\put(155,70){\vector(0,-1){45}}
\put(150,5){$n_3$}
\put(150,190){$m_3$}
\put(225,130){\vector(0,1){45}}
\put(225,180){\circle{5}}
\put(225,20){\circle{5}}
\put(225,70){\vector(0,-1){45}}
\put(220,5){$n_4$}
\put(220,190){$m_4$}
}
\put(0,0){\sf Picture 2. A partial bijection
(matching)
of configurations}
\end{picture}
}

  \begin{center}

{\Large\bf

Spreading maps (polymorphisms),
symmetries of Poisson processes,
and matching summation
}

\bigskip

\sc Yurii A.Neretin

\end{center}

\medskip

{\sc Abstract.}
{\footnotesize
The matrix of a permutation
is a partial case of Markov transition matrices.
In the same way, a measure preserving bijection
of a space $(A,\alpha)$
 with finite measure is a partial case
of Markov transition operators. A Markov transition
operator also can be considered as a map
(polymorphism)
$(A,\alpha)\to (A,\alpha)$,
 which spreads points of $(A,\alpha)$ into
measures on $(A,\alpha)$.

Denote by $\R^*$ the multiplicative group of positive
real numbers and by $\cM$ the semigroup of
measures on $\R^*$.
In this paper, we discuss
 $\R^*$-polymorphisms
 and $\star$-polymorphisms, who are
  analogues of the Markov
 transition operators
 (or polymorphisms) for the groups
 of bijections $(A,\alpha)\to (A,\alpha)$ leaving
  the measure $\alpha$ quasiinvariant;
  two types of the polymorphisms correspond
 to  the cases, when $A$ has finite and
  infinite measure respectively.
  For the case, when the space $A$ itself is finite,
  the $\R^*$-polymorphisms are
  some $\cM$-valued matrices.

  We construct a  functor
  from $\star$-polymorphisms to
  $\R^*$-polymorphisms, it is described
  in terms of summations
  of $\cM$-convolution products
  over matchings
  of Poisson configurations.
 }

 \medskip

{\bf 0.0. Notation and terminology.}
The subject of this paper is  pure measure theory
without any additional structures.

The term "{\it measure}" in this paper means a positive
Borel measure.
The term "subset"
of a space with measure means a Borel measurable subset.

The term {\it space with measure}
means a Lebesgue measure space,
i.e., a space, which is equivalent
to the union of some interval of  $\R$
(the interval can be
finite, infinite or empty)
and some collection
of points having nonzero measures
(this collection can be finite,
countable or empty).
We say that the measure is {\it continuous},
if all points  have zero measure.

We denote spaces with measure
by $(A,\alpha)$, $(B,\beta)$, $(M,\mu)$
etc., the Latin capital letter denotes the space,
the Greek letter denotes the measure.

All our measures are defined on Borel
$\sigma$-algebras.

The symbol
$\R^*$ denotes the multiplicative group
of  positive real numbers.
By $\cM$ we denote the space of finite positive
measures on $\R^*$. We equip
this space with  the weak convergence;
a sequence $\fu_j\in\cM$
weakly converges to $\fu\in\cM$, if for any
bounded continuous function $\psi$
on $\R^*$
we have the convergence
$\int \psi(x)\,d\fu_j(x)\to\int \psi(x)\,d\fu(x)$
(this definition forbid departure of
the measure to $+\infty$ and $0$).
The expression $\mu*\nu$ denotes the convolution
of measures on the multiplicative
group $\R^*$.

\smallskip

{\bf 0.1. Groups.}
We consider 4 groups.
For a space $(A,\alpha)$ with a finite continuous measure,
we define the following groups.

--- $\Ams(A)$  is the group of all measure
    preserving  bijections $A\to A$

     \qquad($\Ams$ is the abbreviation of "automorphisms
    of the measure space"),

--- $\Gms(A)$ is the group of all maps
    $A\to A$
    leaving the measure $\alpha$ quasiinvariant.

For a space $(M,\mu)$ with an infinite continuous measure,
we define two groups:

--- $\Ams_\infty(M)$  is the group of all measure
    preserving  bijections  $M\to M$,

--- $\Gms_\infty(M)$ is the group of all maps $A\to A$
    leaving the measure $\mu$ quasiinvariant
and satisfying the condition
$$
\int_M |q'(m)-1|\,d\mu(m)<\infty
.$$

{\sc Remark.} The group
$\Gms_\infty(M)$ has a homomorphism
to the additive group of $\R$ given by
$$q\mapsto \int_M (q'(m)-1)\,d\mu(m).$$

\smallskip

It turn out to be that all these groups admit
natural embeddings to semigroups
of spreading maps (or polymorphisms).
The semigroup of polymorphisms related
to the group $\Ams(A)$ is a well-known object
(see \cite{Ver}, see also \cite{Kre}, \cite{Nerb}).
Recall its definition.

\smallskip

{\bf 0.2.  The usual polymorphisms.}
Let $(A,\alpha)$, $(B,\beta)$ be spaces with
probability measures.
Consider a probability measure $\fP$ on $A\times B$.
We say that $\fP$ is a
 {\it polymorphism}
 or {\it bistochastic kernel} $\fP:A\to B$
 if

  --- the image of $\fP$ under the
   projection\footnote{it is also called the \it marginal.}
  $A\times B\to A$
   is the measure $\alpha$;

  --- the image of $\fP$ under the
   projection
  $A\times B\to B$
   is the measure $\beta$.

By the Rohlin theorem on conditional
measures (see \cite{Roh}),
 for almost all $a\in A$ there exists
a probability measure $\fP_a$ on $a\times B$
such that
$$
\fP(Q)=\int_A \fP_a(Q\cap \{a\times B\})\,d\alpha(a)
.$$

{\sc Remarks.} 1) Let $U$, $V$ be sets.
Let $R$ be a subset in $U\times V$.
We can consider $R$ as a {\it relation}
or a {\it multivalued map}
$U\to V$. For a point $u\in U$,
its image consists of all the points
$v\in V$ such that $(u,v)\in R$.
For two relations $R\subset U\times V$,
$S\subset V\times W$, we  define
their product $T=SR\subset U\times W$.
It consists of all $(u,w)\in U\times W$ such that there
exists $v\in V$ satisfying the conditions
$(u,v)\in R$, $(v,w)\in S$.
Multivalued maps appear in a natural way
in various branches of mathematics.
The most classical example is the definition
of  algebraic functions $\C\to\C$.
Recall that an algebraic function
 is a subset
in $\C \times \C$ satisfying a polynomial equation
$p(x,y)=0$.

2) Nonformally, a polymorphism
$\fP$ is some kind of a multivalued map
 that
spreads each point $a\in A$ into the measure
$\fP_a$, i.e.
we know not only the image of a point,
but also a
probability distribution on its image.

3) Also polymorphisms are continuous
analogues of  Markov transition
matrices (see \cite{Ver} for detailed
explanations, see also \cite{Hop}).

\smallskip

{\sc Example.} Let $q:A\to A$ be a measure
preserving bijection. Consider its graph
$\graph(q)$, i.e.,
the subset of $A\times A$ consisting of
all the points $(a,q(a))$.
Consider the map $A\to A\times A$
given by $a\mapsto (a,q(a))$.
The image  $\fP_q$ of the measure $\alpha$
with respect to this map
is a measure supported by $\graph(A)$.
Obviously, $\fP_q$
is a polymorphism.

\smallskip

{\sc Example.} The measure $\alpha\times\beta$
is a polymorphism $(A,\alpha)\to (B,\beta)$.
Nonformally, this polymorphism
is the total "uniform spreading"
of the set $A$ along the set $B$.

\smallskip

Let $\fP:(A,\alpha)\to (B,\beta)$
and $\fQ:(B,\beta)\to (C,\gamma)$
be two
polymorphisms. Let $\fP_a(b)$
and $\fQ_b(c)$ be the corresponding systems
of conditional measures. We
define the product
$\fR =\fQ\fP :(A,\alpha)\to(C,\gamma)$
in the terms of
these conditional
measures
\begin{equation}
\fR_a(c)=\int_B \fQ_b(c)\,d\fP_a(b).
\end{equation}

Denote by $\Poll(A,B)$ the set
 of all polymorphisms $A\to B$.

 The set $\Poll(A,A)$ is a semigroup
 with respect to the multiplication.
 This semigroup contains the group
 $\Ams(A)$.

Let $\fP_j,\fP:(A,\alpha)\to (B,\beta)$
be polymorphisms.
We say that the sequence $\fP_j$ converges
to $\fP$ if for each measurable subsets
$U\subset A$, $V\subset B$
the sequence of real
numbers $\fP_j(U\times V)$
converges to $\fP(U\times V)$.

It is readily seen that the space $\Poll(A,B)$
is compact.

It is easy to show (see \cite{Ver}, \cite{Nerb})
 that the group $\Ams(A)$
is dense in the semigroup $\Poll(A,A)$.

\smallskip

{\sc Example.} Let $q\in\Ams(A)$ be a mixing
(i.e., for any subsets $U$, $V\in A$
the measure $\alpha (U\cap q^n(V))$ tends
to $\alpha(U)\times \alpha(V)$ as $n\to +\infty$).
Then $q^n$ converges to the
"uniform spreading" $\alpha\times \alpha$
in $\Poll (A, A)$.
There is a wide literature on
polymorphisms in the ergodic theory,
see \cite{dJR}, \cite{Kre}, \cite{Ver}.

\smallskip

{\sc Remark.} In fact, we have the category of
polymorphisms.
The objects are  Lebesgue spaces
 with probability measure,
 and morphisms $A\to B$ are polymorphisms.
 For groups $\Gms$, $\Ams_\infty$, $\Gms_\infty$,
  we also
 describe below some categories, whose objects
 are Lebesgue spaces with measure.

\smallskip

{\bf 0.3. Closure of an
invariant action and the extension
problem.}
Consider a group $G$ acting by measure
preserving maps on a space $A$ with
a finite continuous measure $\alpha$.

\smallskip

{\sc Extension problem.}
 {\it For a given action of a
  group $G$, to find
the closure $\Gamma$
of $G\subset \Ams(A)$  in the semigroup
of polymorphisms of $A$.}

\smallskip

It seems that nothing interesting
can happen for connected non-Abelian
 Lie groups $G$
(the case of Abelian groups is another story).
Nevertheless, the problem becomes
very nontrivial for infinite-dimensional
("large") groups
\footnote{It seems
that the term "large" group
introduced by Vershik  is better
than "infinite dimensional" group. For instance,
our groups $\Ams$, $\Ams_\infty$,
 $\Gms$, $\Gms_\infty$ have no structure
 of a manifold, but they are "very large".}.
 Indeed, the semigroup $\Pol(A,A)$
 is compact, and hence the semigroup
 $\Gamma$
 also is compact. Obviously,
 any compactification
 of a large group $G$ essentially differs
 from the group $G$ itself.

 \smallskip

 {\bf 0.4. Another variant of extension
 problem.}
In many cases, the semigroup $\Gamma$
is known by {\tt a priory} reasons.
Assume that $G$ has some collection
of unitary representations.
Then usually {\it there exists a canonical
semigroup $\Gamma\supset G$
such that any
unitary representation of the group $G$
admits a canonical extension to a representation
of the semigroup $\Gamma$.}
This statement was claimed by G.I.Olshanski
in the end of 70-ies
(see \cite{Ols1}--\cite{Ols2}, \cite{Ners}),
for more details see \cite{Nerb}).

This is not a general theorem
but an experimental  fact.
Nevertheless, in the most
 cases,
there exists a constructive description of
the semigroup $\Gamma$ and its representations,
see \cite{Nerb}.

For many groups $G$,
there exist also {\tt a priory}
theorems about
the extension of representations
to $\Gamma$.

\smallskip

{\sc Examples.}
1) For $G=\Ams(A)$,
the semigroup
$\Gamma$ is the semigroup
$\Poll(A,A)$. The {\tt a priory}
 theorem on extension of representations
is obtained in \cite{Nerbi}, see also \cite{Nerb},
Section 8.4.

2) For $G=\Ams_\infty$, $\Gms$, $\Gms_\infty$,
the semigroups $\Gamma$ are the semigroups
of polymorphisms defined below (Sections 1--2),
see \cite{Nerbi}.

3) If $G$ is the complete orthogonal
group of a Hilbert space, then
the semigroup $\Gamma$
is the semigroup $Contr$
of all operators in the real Hilbert space
with the norm $\le 1$, \cite{Ols1}.

4)
 More interesting examples with inordinate $\Gamma$
  are contained
in \cite{Ols2}, \cite{Ners}, \cite{Nerb}.

\smallskip

In many cases (see \cite{Nerd}), it can be easily
shown, that any homomorphism
$G\to \Ams(A)$ can be extended to
a homomorphism
$\Gamma\to\Poll(A,A)$.

\smallskip

Thus we obtain the following variant
of the extension problem
(this variant is not exactly equivalent to previous
one).

{\it Consider any case, then $\Gamma$ is known.
 For a given measure preserving action
of a "large" group $G$, to find
 an explicit description of
the homomorphism
$\Gamma\to \Poll$.}

{\bf 0.5. The purpose of the paper.}
I know only one work that can
be attributed to this extension
problem.
Consider the well-known action of
the complete infinite dimensional
orthogonal group $O(\infty)$
on the space with Gauss measure (see \cite{Seg},
\cite{Shi},
see also \cite{Nerb}).
The corresponding homomorphism
of the semigroup of contractions
$Contr$ to $\Poll$
was explicitly described by Nelson \cite{Nel}.

A few interesting measure preserving
actions of large groups are known,
and hence the polymorphism extension problem
has a restricted interest.
But the zoo of quasiinvariant
actions is very rich (see survey \cite{Nerd}
and  recent papers
on virtual permutations and
Pickrell' type inverse
limits of symmetric spaces
\cite{Pic}, \cite{Ker}--\cite{KOV},
\cite{BO1}--\cite{BO2}, \cite{Nerh}).

It turn out to be that there are polymorphism-like
semigroups related to all the groups
$\Ams_\infty$, $\Gms$,  $\Gms_\infty$.
We describe  them explicitly below in Sections 1-2.

It seems that the most important of these objects
 is the semigroup
$\Pol(A,A)$ related to
the group $\Gms(A)$, its elements
are measures on
$$A\times A\times \R^*$$
satisfying some additional conditions.
These $\R^*$-polymorphisms
 can be considered as  "spreading maps",
but they spread not only points; also
 Radon--Nykodim
derivatives at points are spreaded.

For each quasiinvariant action
of a large group $G$ on a measure space
$(A,\alpha)$, we obtain
a problem about extension
of the
 homomorphism
$G\to\Gms(A)$ to the homomorphism
from $\Gamma$ to $\Pol(A)$.

The purpose of this paper
is to understand the degree
of the interest of this problem.
We consider the simplest (for my test)
nontrivial quasiinvariant action of
a large
group on a measure space (see the next
subsection).

\smallskip

{\bf 0.6. Poisson configurations.}
Let $M$ be a space with a continuous infinite
 measure
 $\mu$. Denote by $\Omega(M)$
the space of all countable subsets
$\bm=(m_1,m_2\dots)$ in $M$.
We define the {\it Poisson measure} $\nu$
on $\Omega(M)$ by the following conditions.

 $1^*$. Let $A\subset M$ have finite measure.
 Denote by ${\cal S}_k(A)$ the set of
 all $\bm\in \Omega(M)$ such that
the set $A\cap \bm$ consists of $k$ points.
Then
$$\nu({\cal S}_k(A))=
\frac {\mu(A)^k}{k!} e^{-\mu(A)}.$$

 $2^*$. Let sets $A_1$, \dots, $A_n$
 be mutually disjoint. Then
 the events
${\cal S}_{k_1}(A_1)$,\dots, ${\cal S}_{k_n}(A_n)$
are independent, i.e.,
$$
\nu\bigl(
\bigcap\limits_{j=1}^n{\cal S}_{k_j}(A_j)\bigr)=
\prod_{j=1}^n \nu\bigl({\cal S}_{k_j}(A_j)\bigr)
.$$

 It is easily shown that these conditions
 define a unique probability
measure on $\Omega(M)$.


{\sc Theorem.}
{\it The measure $\nu$ on $\Omega(M)$
is quasiinvariant with respect to
the group
$\Gms_\infty(M)$,
the Radon--Nykodim derivative
of the transformation
\begin{equation}
\bm=(m_1,m_2,\dots)\mapsto q\bm=(qm_1,qm_2,\dots)
,\qquad q\in\Gms_\infty(M)
,\end{equation}
 is given by the formula}
\begin{equation}
\exp\Bigl\{-\int_M(q'(m)-1)\,d\mu(m)\Bigr\}
\prod_{m_j\in\bm}
q'(m_j)
.\end{equation}

This quasiinvariance was obtained by
Vershik, Gelfand, Graev \cite{VGG}
(in their paper, $q$ was
a finitely supported
diffeomorphism of a manifold),
the infinitesimal version of Theorem
0.1 was obtained earlier by Goldin, Grodnik,
Powers, Sharp, Menikoff
\cite{GGP}, \cite{Men} (see also \cite{Kon});
the
variant of Theorem given above was obtained
in \cite{Nerbi}, for details see  \cite{Nerb},
Section X.4.
Spherical functions on the group
$\Gms_\infty$ with respect to the group
$\Ams_\infty$
 are discussed
in \cite{Ism}.

{\bf 0.7. The result of the paper.}
Thus we have the canonical homomorphism
\begin{equation}
\Gms_\infty(M)\to\Gms\bigl(\Omega(M)\bigr).
\end{equation}

In this paper, we describe explicitly
the  homomorphism
of the semigroups of
polymorphisms extending
(4).

In fact, we construct some canonical  family
of measures ($\R^*$-polymorphisms) on
$$\Omega(M)\times\Omega(M)\times \R^*.$$
They can be interpreted as 'spreading maps'
of the space $\Omega(M)$.
Any such 'map' can be obtained as a limit
of the transformations (2); thus our $\R^*$-polymorphisms
themself are some kind of symmetries
of Poisson processes.
We define our $\R^*$-polymorphisms
of $\Omega(M)$ in the terms of the
{\it matching summation}
formula (18).
In fact, this formula is similar to the expressions
for the Taylor coefficients of
$$
\sum\sigma_{kl}z^k u^l=
\exp\Bigl\{\sum_{k,l} a_{kl} z_k u_l +
\sum_k b_k z_k + \sum_l c_l u_l +d\Bigr\}.
$$
In these expressions, the scalars
 $a_{kl}$, $b_k$, $c_l$, $d$
are replaced by measures on $\R^*$
and
the products of scalars
 are replaced by convolutions of the measures.
 The analogue of $\exp(d)$
 in the formula (18) is a sophisticated
expression.

Matching summation  itself
appears in mathematics in various situations (see
\cite{LP}, \cite{Nerp}),
but such combinatorial expressions with
measures seem unusual.


 This work is a continuation
of  \cite{Nerp}, but logically
these two papers are independent.

\smallskip

{\bf 0.8. Structure of the paper.}
 Section 1 contains preliminaries
 on $\R^*$-polymorphisms, i.e., polymorphisms
 related to the group $\Gms$.
In Section 2, we define
$\star$-polymorphisms
 related to the group $\Gms_\infty$.

In Section 3, for any  $\star$-polymorphism,
 we construct
an $\R^*$-polymorphism
of the corresponding spaces
of Poisson configurations.

The result of this paper
is the formula (18) and Theorems A-B.

\smallskip

{\bf Acknowledgments.}
I thank A.M.Vershik for
explanations of $\Poll(\cdot)$
 and discussions of
polymorphisms.
I thank the administrators of
Erwin Schr\"odinger Institute (Vienna)
for hospitality.

\medskip

{\large\bf 1. $\R^*$-polymorphisms}

\medskip

In Sections 1--2, we apply
the double coset
multiplication machinery  for producing
the semigroups of polymorphisms.
In fact, we also give motivation
independent definitions
of $\R^*$-polymorphisms and
$\star$-polymorphisms in 1.5-1.8
and 2.7-2.8.
But it seems that the double coset
motivation is
really necessary in Section 2.


On double coset multiplication
and similar operations,
see \cite{Bro},
\cite{Ver}, \cite{Ols1}--\cite{Ols2},
a relatively complete
list of such constructions
is contained in the book \cite{Nerb},
its Russian edition is more complete.

\smallskip

Consider a group $G$ and  subgroups $H$,
$K$.
The {\it double coset
space} $H\setminus G/K$ is a
quotient space of $G$ with respect
to the equivalence
 relation
$$
g\sim k g h,\qquad\text{where}\quad g\in G, h\in H,
k\in K
.$$

 The equivalence classes
are called {\it double cosets}.

\smallskip

{\bf 1.1. Double coset multiplication on
$\Ams\setminus\Gms/\Ams$.}
Fix a space $(A,\alpha)$ with a continuous
 probability measure.
Let
$g\in \Gms(A)$.
Consider the map $A\to \R^*$
given by
$$ a\mapsto g'(a).$$
Denote by $\fu_g$ the image of this map.
Obviously, $\fu_g$ is a probability measure
on $\R^*$, this measure also satisfies
the condition
\begin{equation}
\int_{\R^*} x\, d\fu(x)=1
.\end{equation}
The last property
 is equivalent to
 $$\int_A g'(a)\,d\alpha(a)=1.$$
Denote by $\cL$ the set of all
probability measures on $\R^*$
satisfying the condition (5).

Obviously, for any
$h_1, h_2\in\Ams(A)$, we have
$$
\fu_{h_1gh_2}=\fu_g
,$$
i.e., the map $g\mapsto \fu_g$
is constant on double cosets.
It is readily seen that the map
$$\Ams(A)\setminus\Gms(A)/\Ams(A)\to \cL $$
defined by by $g\mapsto \fu_g$
is a bijection.

We claim that there
exists
 a natural multiplication
on the double coset space $\Ams\setminus\Gms/\Ams$.

Consider $\fv,\fw\in\cL$.
Consider the representatives
$p,q$ of the corresponding
double cosets, i.e., $\fu_p=\fv$,
$\fu_q=\fw$. Of course,
the
element $\fu_{pq}$
depends on the choice of $p$ and $q$
(and it is not determined by $\fu$, $\fw$).

Nevertheless, there exists the following
nonformal reasoning.
Let $h\in\Ams(A)$ be "as general as possible".
It is clear that $h$ "very strongly mix"
the space $A$,
this imply that $\fu_{phg}$
is very close to the convolution
$\fu_p*\fu_q$.
For an 'absolutely generic'
$h$, we will obtain the convolution $\fu_p*\fu_q$
itself. Thus the multiplication
of double cosets is the convolution
of the corresponding measures $\fu_q$.

One of ways to say the same
reasoning carefully
is the following.

We say that a sequence $h_n\in \Ams(A)$
is {\it generic}
if it converges to the uniform spreading
(see 0.3) in $\Poll(A,A)$.
The following is a rephrasing of the definition:
a sequence $h_j$ is generic
if:
$$\forall B,C\subset A \qquad
\lim\limits_{n\to\infty}
  \alpha\bigl(h_n(B)\cap C\bigr)
  =\alpha(B)\alpha(C).$$

{\sc Remark.}
  If $A$ is a space with finite nonprobability
  measure, then the definition of
 a {\it generic} sequence $h_n$ has the form
$$\qquad\qquad\qquad
\forall B,C\subset A \qquad
\lim\limits_{n\to\infty}
  \alpha\bigl(h_n(B)\cap C\bigr)
  =\frac{\alpha(B)\alpha(C)}{\alpha(A)^2}.
\qquad\qquad (5.a)$$

The following statement is obvious.

\smallskip

{\sc Lemma.}
{\it For a generic sequence $h_n$ and any
$p, q\in\Gms(A)$,
the sequence $\fu_{ph_nq}$
weakly converges to
$\fu_p *\fu_q$.}

\smallskip

Thus we define the multiplication
of the double cosets as the convolution
of the corresponding measures.

\smallskip

{\bf 1.2. Partitions.}
Let $A$ be a  space  with a
probability measure.
Consider its finite or countable partition
$$T:A=A_1\cup A_2\cup\dots $$
By $A/T$ we denote the quotient-space, i.e.,
the countable space, where
the measures of the  points are
$\alpha(A_1)$,  $\alpha(A_2)$,\dots.
Denote by $\Ams(A\bigl| T)$
the group
$$
\Ams(A\bigl| T)=\Ams(A_1)\times \Ams(A_2)\times\dots
\subset\Ams(A).$$

{\bf 1.3. Double cosets.}
Consider a space $(A,\alpha)$
 with a continuous measure.
Consider two partitions
of $A$
(they can coincide)
$$S:A=A_1\cup A_2\cup\dots;
\qquad T:A=B_1\cup B_2\cup\dots $$
Consider the quotients
$A/S$ and $A/T$. Denote their
points by $a_1$, $a_2$, {\dots} and
$b_1$, $b_2$, {\dots} respectively.
Denote the measures of the points
by
$\alpha_1$, $\alpha_2$, {\dots}
and $\beta_1$, $\beta_2$, \dots.

Consider the double cosets
\begin{equation}
\Ams(A\bigl| S)\setminus \Gms(A)/  \Ams(A\bigl|T)
.\end{equation}
Fix $p\in\Gms(A)$.
For each pair $A_i$, $B_j$, consider
the set
$A_i\cap p^{-1}(B_j)$. Denote by
$\fp_{ij}$ the image of the measure
$\alpha$ restricted to $A_i\cap p^{-1}(B_j)$
under the map
$$A_i\cap p^{-1}(B_j)\to \R^*.$$
Thus we obtain an $\cM$-valued matrix
\begin{equation}
\fP=\begin{pmatrix}
\fp_{11}&\fp_{12}&\dots\\
\fp_{21}&\fp_{22}&\dots\\
\vdots &\vdots &\ddots
\end{pmatrix}
,\end{equation}
where each $\fp_{ij}$ is a measure on $\R^*$;
these measures satisfy the conditions
\begin{gather}
\sum_i\int_{\R^*} x d\fp_{ij}(x)=\beta_j,\\
\sum_j\int_{\R^*}  d\fp_{ij}(x)=\alpha_i
.\end{gather}
The origin of these conditions are the identities
\begin{gather*}
\alpha(A_i)=\sum_j\alpha\bigl(A_i\cap p^{-1}(B_j)\bigr);
\\
  \alpha(B_j)=\sum_i\alpha\bigl(p(A_i)\cap B_j\bigr)
=\sum_i
\int_{A_i\cap p^{-1}(B_j)}p'(a)\, d\alpha(a)
.\end{gather*}
It is readily seen that the map
$p\mapsto\fP$ induces a bijection
from the double coset space (6)
to the space of all matrices (7)
satisfying the conditions (8)--(9).

\smallskip

{\bf 1.4. The multiplication of double cosets.}
For a space $(A,\alpha)$ with a continuous
probability measure, consider 3 partitions
(they can coincide)
$$S:A=A_1\cup A_2\cup\dots;
\qquad T:A=B_1\cup B_2\cup\dots;
\qquad R:A=C_1\cup C_2\cup\dots $$
Denote by $\beta_1$, $\beta_2$,\dots
the measures of the sets $B_1$, $B_2$, \dots.

We intend to define the multiplication
of double cosets
\begin{multline*}
\Ams(A\bigl| S)\setminus \Gms(A)/ \Ams(A\bigl| T)\quad
\times\quad
\Ams(A\bigl| T)\setminus \Gms(A)/ \Ams(A\bigl| R)
\quad \to \\ \to\quad
\Ams(A\bigl| S)\setminus \Gms(A)/ \Ams(A\bigl| R)
.\end{multline*}
We say that a sequence
$$h_n=(h_n^{(1)},h_n^{(2)},\dots)\in\Ams(A\bigl| T)=
\prod_j \Ams(A_j)$$
is {\it generic} if for each $j$ the sequence
$h_n^{(j)}$ is generic in $\Ams(A_j)$.

Consider a transformation
$p\in\Gms(A)$ and the corresponding double coset
in
$\Ams(A\bigl| S)\setminus \Gms(A)/ \Ams(A\bigl| T)$,
i.e., consider
 the matrix $\fP=\{\fp_{ij}\}$.
Consider a transformation
$q\in \Gms(A)$
and consider the corresponding double coset
in
$\Ams(A\bigl| T)\setminus \Gms(A)/ \Ams(A\bigl| R)$.
Denote by $\fQ=\{\fq_{jk}\}$
the corresponding matrix.

For the product $qh_np$
denote by $\fR_n$ the
corresponding double coset in
$\Ams(A\bigl| S)\setminus \Gms(A)/ \Ams(A\bigl| R)$.

 {\sc Lemma.} {\it The sequence
of $\cM$-valued matrices $\fR_n$ converges
{\rm(}elementwise{\rm)} to the matrix

\begin{equation}
\fR=\begin{pmatrix}
\fr_{11}&\fr_{12}&\dots\\
\fr_{21}&\fr_{22}&\dots\\
\vdots &\vdots &\ddots
\end{pmatrix}
=\begin{pmatrix}
\fp_{11}&\fp_{12}&\dots\\
\fp_{21}&\fp_{22}&\dots\\
\vdots &\vdots &\ddots
\end{pmatrix}
\begin{pmatrix}
\beta_1^{-1} & 0&\dots\\
0& \beta_2^{-1}& \dots\\
\vdots&\vdots&\ddots
\end{pmatrix}
\begin{pmatrix}
\fq_{11}&\fq_{12}&\dots\\
\fq_{21}&\fq_{22}&\dots\\
\vdots &\vdots &\ddots
\end{pmatrix}
,\end{equation}
where
the product of matrix elements is
the convolution
of measures on $\R^*$, i.e.,
$$
\fr_{ik}=\sum_j \frac 1{\beta_j}
  \fp_{ij}*\fp_{jk}
.$$}

Formula (10) defines the required product of
the double cosets.

Now we will make a definition from
this Lemma. The definition is formal
and motivation independent.

\smallskip

{\bf 1.5. $\R^*$-polymorphisms
of countable spaces.}
Consider a countable (or finite)
 space $A$ with a probability
measure. We denote its points by
$a_1$, $a_2$, \dots,
we denote their measures by $\alpha_1$,
$\alpha_2$, \dots.
Let $A$, $B$ be two countable spaces.
Then an $\R^*$-polymorphism $A\to B$
is an $\cM$-valued matrix (7)
satisfying the conditions (8)--(9).

Let $A$, $B$, $C$ be countable (or finite) spaces
with probability measures.
Let $\fP:A\to B$, $\fQ:B\to C$ be
$\R^*$-polymorphisms. Then their product
$\fR=\fQ\fP:A\to C$ is defined by
the formula (10).

\smallskip

{\bf 1.6. $\R^*$-polymorphisms
in general case,}
(\cite{Nerbi}, \cite{Nerb}).
Consider spaces
$(A,\alpha)$, $(B,\beta)$  with
probability measures.
An $\R^*$-polymorphism
$\fP:A\to B$ is a measure
$\fP$ on $A\times B\times\R^*$
satisfying two conditions

1. The image of the measure $\fP$
under the projection
$A\times B\times\R^*\to A$ is $\alpha$.

2. Denote by $x$ the coordinate on $\R^*$.
Consider the measure $x\cdot \fP$.
We require the image of $x\cdot \fP$
under the projection $A\times B\times\R^*\to B$
to be
$\beta$.

\smallskip

Denote by $\Pol(A,B)$ the set
of all $\R^*$-polymorphisms $A\to B$.

\smallskip

{\sc Example.} Consider
a space $A$ with a continuous probability
measure.
Consider $q\in\Gms(A)$. Denote by $q'(a)$
its Radon--Nykodim derivative.
Consider the map $A\to A\times A\times\R^*$
given by
$$a\mapsto (a, q(a),q'(a)).$$
Denote by $\fP(q)$ the image
of the measure $\alpha$ under this map.
Then $\fP(q)$ is an element
of $\Pol(A,A)$.

\smallskip

{\bf 1.7. Convergence.}
Consider general spaces $(A,\alpha)$,
$(B,\beta)$ with
probability measures.
Consider an arbitrary
 $\R^*$-polymorphism $\fP\in\Pol(A,B)$.
Fix Borel subsets $M\subset A$, $N\subset B$
and
 consider the restriction of the measure
$\fP$ to $M\times N\times \R^*$.
Denote by $\fp[M,N]$ the image of this restriction
under the map $M\times N\times \R^*\to \R^*$.

We say that the sequence $\fP_j\in\Pol(A,B)$
converges to $\fP\in\Pol(A,B)$
if
for each $M\subset A$, $N\subset B$,

1. the
 sequence $\fp_j[M,N]$ weakly converges
to $\fp[M,N]$

2. the
 sequence $x\fp_j[M,N]$ weakly converges to $x\fp[M,N]$

 \smallskip

 See examples of the convergence below in 1.11.

 \smallskip

 {\sc Remark.} Consider a space $A$
 with a continuous probability measure $\mu$.
 It is easy to prove that
 the group $\Gms(A)$
 is dense in the semigroup
$\Pol(A,A)$ (\cite{Nerbi}).

\smallskip

{\bf 1.8. Definition
of product of $\R^*$-polymorphisms
 in general case.}
Let $A$ be a  space  with a
probability measure.
Consider its finite or countable partition
$$T:A=A_1\cup A_2\cup\dots $$
By $A/T$ we denote the quotient-space, i.e.,
the countable space, where
the measures of  points are
$\alpha(A_1)$,  $\alpha(A_2)$,\dots.

Consider also a partition of
a space $B$
$$S: B=B_1\cup B_2\cup\dots $$
For any  $\fP\in \Pol(A,B)$,
we define
$$\fP_{T,S}^\downarrow\in \Pol(A/T, B/S)$$
as the matrix consisting of the measures
$\fp[A_i,B_j]$ (see 1.7).

Conversely, consider
$$\fR\in \Pol(A/T, B/S).$$
This is a matrix, whose matrix elements
$\fr_{ij}$ are measures on $\R^*$.
For each $A_i$, $B_j$, consider the
measure on $A_i\times B_j\times \R^*$
given by
$$\frac{\alpha}{\alpha(A_i)}
\times\frac{\beta}{\beta(B_j)}\times\fr_{ij}.$$
This defines some measure $\fR_{T,S}^\uparrow$ on
$$A\times B\times\R^*=
\bigcup\limits_{ij}A_i\times B_j\times \R^*
.$$
Obviously,
$$\fR_{T,S}^\uparrow\in \Pol(A,B). $$

Also, $\bigl(\fP_{T,S}^\uparrow\bigr)^\downarrow=\fP$,
and, obviously, $\bigl(\fR_{T,S}^\downarrow\bigr)^\uparrow$
is not $\fR$.

We say that a sequence $T^{(j)}$
of partitions is {\it approximative}, if
a partition $T^{(j+1)}$ is a refinement
of $T^{(j)}$ and elements of
the partitions
generate the Borel $\sigma$-algebra
 of $A$.

Now we are ready to define
the product of $\fP\in\Pol(A,B)$
and $\fQ\in\Pol(B,C)$.
Consider approximative sequences
of partitions $T^{(j)}$, $S^{(j)}$,
$U^{(j)}$ of the spaces $A$, $B$, $C$.
We define the polymorphism
$\fR=\fQ\fP\in\Pol(A,C)$ as
$$
\lim_{j\to\infty}
\Bigl(
 \fQ_{S^{(j)},U^{(j)}}^\downarrow
 \fP_{T^{(j)},S^{(j)}}^\downarrow
 \Bigr)_{T^{(j)},U^{(j)}}^\uparrow
.$$

{\sc Remark.} For any group $G$ it is possible
to define $G$-polymorphisms in the same way,
see \cite{Nerbi}, \cite{Nerb}.
For some groups $G$, there exist nontrivial functors
from category of polymorphisms to the category
of Hilbert spaces and operators
(\cite{Nerbi}, \cite{Nerb})
($G=SL_2(\R)$, $O(1,n)$, $U(1,n)$,
these functors extend the so-called
 Araki multiplicative
integral construction, see \cite{Ara}, \cite{VGG1});
 for some groups
$G$ there exist nontrivial central extensions
of  categories of $G$-polymorphisms
(for $G=Sp(2n,\R)$, $U(p,q)$, $SO^*(2n)$,
\cite{Nern}).

 {\sc Remark.} It seems that some polymorphism-like
structures appears in the mathematical
hydrodynamics, see \cite{Bre}.

\smallskip

{\bf 1.9. Remark. Action of $\R^*$-polymorphisms
on spaces $L^p$.}
Let $w=u+iv$ be in $\C$, let
$0\le u\le 1$.
Let $A$ be a space with a continuous probability measure.
The group $\Gms(A)$ acts in the space
$L^{1/u}(A)$ by the isometries
$$T_w(q)f(a)=f(q(a))q'(a)^w$$
Let us extend this action to the action
of $\R^*$-polymorphisms.

Let $(A,\alpha)$, $(B,\beta)$
be  spaces with probability
measures.

{\sc Proposition.}
 {\it Let $\fP\in\Pol(A,B)$.
Then the expression
$$S_w(\fP|f,g)=
\iiint_{A\times B\times \R^*}
f(a)g(b)x^{u+iv}\,d\fP(a,b,x)
$$
is a bounded bilinear form on
 $L^{1/(1-u)}(A)\times L^{1/u}(B)$
and moreover}
$$
|S_w(\fP|f,g)|\le \|f\|_{L^{1/(1-u)}}\cdot
\|g\|_{L^{1/u}}
.$$

Let us define the linear operator
$$T_w(\fP):L^{1/u}(B)\to L^{1/u}(A)$$
by
 the duality condition
 $$
 \int_A f(a)T_w(\fP) g(a)\,d\alpha(a)
 =S_w(\fP|f,g)
 .$$
 Obviously,
 $$\|T_w(\fP)\|\le 1.$$

 {\sc Proposition.}
 {\it For each spaces $A$, $B$, $C$
 with probability measures
 and each $\fP\in\Pol(A,B)$, $\fQ\in\Pol(B,C)$,}
 $$T_w(\fQ) T_w(\fP)=T_w(\fQ\fP).$$

{\bf 1.10. Remark. Action of $\R^*$-polymorphisms
on $\cM$-valued functions.}
Let $(A,\alpha)$ be a space with a probability measure.
Denote by $\cS(A)$ the space of all  functions
$a\mapsto\nu_a$ on
$A$ taking values in $\cM$
satisfying the condition
$$\int_{A}\int_{\R^*}
x \,d\nu_a(x)\,d\alpha(a)=1
.$$
Denote by\, $\bullet$ \,the single-point space with
a probability measure.
The space $\Pol(A,\bullet)$
is identified in the obvious way
with the
space $\cS(A)$.

Any element of $\Pol(B,A)$ induces
the map $\Pol(A,\bullet)$ to $\Pol(B,\bullet)$
given by the formula
$$\fU\mapsto \fU\fP;\qquad \fU\in\Pol(A,\bullet).$$
Thus we obtain the canonical map
$$\Theta_\fP:\cS(A)\to \cS(B).$$
Obviously, for any $\fP\in\Pol(B,A)$,
$\fQ\in\Pol(C,B)$,
we have
$$
\Theta_{\fP}\Theta_{\fQ}=\Theta_{\fP\fQ}
.$$

{\bf 1.11. Remarks.  Examples of convergence.}
1)
Let $A=B$ be the interval $[0,1]$.
Consider the sequence $q_n$ of monotonic maps
$[0,1]\to[0,1]$ given by
$$
q_n(a)=a+\frac{1}{2\pi n}\sin(2\pi n a)
.$$
Then the limit $\fP$ of $q_n$ is a measure
on $[0,1]\times[0,1]\times\R^*$
supported by the set consisting of the points
$$ (a,a, x); \qquad 0<x<2$$
and the density of $\fP$ on this set is given by
$$
\frac{da\,dx}{\pi \sqrt{2x-x^2}}
.$$

2) Let $A=B$ be the same.
Then the sequence
$q_n(a)=a^n$ has no limit in $\R^*$-polymorphisms.

3) Let $A,B$ be  spaces with  continuous measures.
Let $S, T$ be
their partitions. Let $g_n\in\Ams(A\bigl|S)$,
$B_n\in\Ams(A\bigl|T)$,
be generic sequences.
Then
$$\lim\limits_{n\to\infty}\Big\{
\lim\limits_{m\to\infty}
h_n\fP g_m\Big\}=\fP^\downarrow_{S,T}
$$

4)  Let $A,B$ be  spaces with  continuous measures.
Let $S^{(n)}, T^{(n)}$ be
approximative sequences of
their  partitions. Then, for any $\fP\in\Pol(A,B)$,
$$
\lim\limits_{n\to\infty}
 \bigl(\fP^\downarrow_{S^{(n)}, T^{(n)}}\bigr)%
    ^\uparrow_{S^{(n)}, T^{(n)}}=\fP.
    $$

{\bf 1.12. Remark.
How to formulate  problem of limit
behavior of powers of a polymorphism?}
For  $\fP\in\Pol(A,A)$,
denote by $\fP^n$ its powers.
If $\fP\in\Ams(A)\subset\Poll(A,A)$,
 then  the problem of limit behavior of the
 powers is the problem of the ergodic theory.
 If $A=\bullet$ is a single-point set,
 then the limit behavior of $\fP^n$
 is described by the central limit
 theorem.
 The following problem is an attempt
 to unite the both  subjects
 of the classical theories mentioned above.

We notice that the group
$\R^*$ admits an one-parametric family
of automorphisms $x\mapsto x^\alpha$,
there $\alpha\in\R\setminus 0$. These automorphisms
induce the one parametric family of automorphisms
of the semigroup $\cM$, i.e.
$$\fu(x)\mapsto\fu(x^\alpha),\qquad\fu\in\cM.$$
The last automorphisms induce
automorphisms of the semigroup
of {\it all}
$\cM$-valued $n\times n$ matrices (7)
equipped with the multiplication (10)%
\footnote{These automorphisms break the condition (8).
But the product (10) itself exists without the
conditions (8)--(9)}.

For any $\fP\in\Pol(A,A)$,
and any $\alpha\in\R\setminus 0$
we define the measures
$\fP(a,b,x^\alpha)$ on $A\times A\times\R^*$
as the image of $\fP$ under the map
$(a,b,x)\mapsto (a,b,x^\alpha)$.

We obtain the following problem:
 {\it Is it possible to find a sequence
 $\alpha_n$ such that
the sequence $\fP(a,b,x^{\alpha_n})^n$
converges to some nontrivial limit?}

\bigskip

{\large\bf 2. Polymorphisms of bordered spaces}

\bigskip

{\bf 2.1. The classes $\cM^\tr$, $\cM^\chtr$
of measures on $\R^*$.}
Let $\fu$
be a measure on $\R^*$.
We say that $\fu$ belongs to
the class $\cM^\tr$,
if
$$
\int_{\R^*}\,d\fu(x)<\infty,\qquad
\int_{\R^*}x\,d\fu(x)<\infty
.$$
We say that $\fu$ is an element
of the class $\cM^\chtr$,
if
$$\int |x-1|\, d\fu(x)<\infty.$$
{\it For the class $\cM^\chtr$, we admit
infinite atomic measures supported by
$x=1$}.

\smallskip

We also define the convergence in
 $\cM^\tr$ and $\cM^\chtr$.
A sequence $\fu_j$ converges to $\fu$ in $\cM^\tr$
if $\fu_j$ weakly converges to $\fu$
and $x\fu_j$ weakly converges to $x\fu$.
A sequence
 $\fu_j$ converges to $\fu$ in $\cM^\chtr$
if $|x-1|\fu_j$ weakly converges to $|x-1| \fu$.

{\bf 2.2. Bordered spaces.}
Let $M$ be a Lebesgue measure space
(we admit the case, when $M$ is an empty set).
We define the
corresponding {\it bordered space}
$$M^\star=M\cup \xi_\infty^M,$$
where $\xi_M$ is a formal point.

\smallskip

{\sc Remark.}
It is natural to think that
$\xi_\infty=\xi_\infty^M$
is "the point of $M$ at infinity".
Also, it is natural to think,
that the measure of the point
$\xi_\infty$ is $\infty$.

\smallskip

We also define {\it measurable subsets}
in  $\cM^\star$. Let $A\subset M$ be a
Borel subset.

a) The set $A\cup\xi_\infty$
is measurable in $\cM^\star$.

b) If $A$ has a finite measure,
then $A$   is measurable in $\cM^\star$.

All other subsets in $M^\star$
are not measurable.

\smallskip

{\bf 2.3. Partitions of bordered spaces.}
Consider a bordered space
$M^\star$.
Its {\it good  partition $\cU$}
(we will omit the word "good") is
a partition
$$\cU:M^\star=M_1\cup\dots\cup M_k\cup M_\infty$$
into mutually disjoint
subsets
such that $M_1,\dots, M_k$
have finite measure and $\xi_\infty\in M_\infty$.
We say that $M_\infty$
is an infinite element
of the partition, all other elements are {\it finite.}

For the partition $\cU$,
we define the quotient space
 $M^\star /\cU$. It
consists of the  points with the measures
$\mu(M_1)$, \dots,  $\mu(M_k)$
and the point $\xi_\infty$.

For the partition $\cU$, we define the group
of automorphisms of the partition
$$\Ams_\infty(M^\star\bigl|\cU)=
\Ams(M_1)\times \dots\times\Ams(M_k)
\times \Ams_\infty(M_\infty).$$

\smallskip

{\bf 2.4. Multiplication of double cosets
 $\Ams_\infty\setminus \Gms_\infty/\Ams_\infty$.}
For $q\in\Gms_\infty(M)$ we consider
the image $\fu_q$ of the measure
$\mu$ under the map
$M\to\R^*$ given by
$m\mapsto q'(m)$.
Obviously,
$\fu_q\in\cM^\chtr$,
and the map $q\mapsto \fu_q$
defines a bijection
$$
\Ams_\infty(M)\setminus \Gms_\infty(M)/\Ams_\infty(M)
\quad \to\quad \cM^\chtr
.$$
We say that a sequence
$h_n\in\Ams_\infty(M)$
is {\it generic} if for any subsets $A, B\subset M$
having finite measures,
we have
$$
\lim\limits_{n\to \infty}
\mu\bigl(h_n(A)\cap B\bigr)=0
.$$

{\sc Remark.} See formula (5.a).

\smallskip

{\sc Example.} Let $M=\R$. We can give $h_n(x)=x+n$.

\smallskip

Fix a generic sequence $h_n$.
Consider $\fv,\fw\in \cM^\chtr$.
Consider $p,q\in\Gms_\infty(M)$ lying in the corresponding
double cosets.

\smallskip

{\sc Lemma.} {\it Denote by $\fu_n$ the element
of $\cM^\chtr$ corresponding $qh_n p$.
Then}
$$\lim\limits_{n\to \infty}
\fu_n=\fv+\fw
.$$

{\sc Remark.} Compare with Subsection 1.1.

\smallskip

{\bf 2.5. Double cosets.} Let $(M,\mu)$ be a space with
a continuous infinite measure.
Fix two partitions
$$\cU: M=M_1\cup\dots\cup  M_s\cup M_\infty,\qquad
  \cV: M=N_1\cup\dots\cup
  N_t\cup N_\infty
  $$
  of $M$.
  Denote the measures of the sets
  $M_i$ by $\mu_i$ and the measures of $N_j$ by $\nu_j$.
 For any $p\in\Gms_\infty(M)$ and any
 $\alpha=1,\dots,s,\infty$ and
 $\beta=1,\dots,t,\infty$, we define
 the measure $\fp_{\alpha\beta}$
 on $\R^*$ as the image of the measure $\mu$
 under the map
 $$M_\alpha\cap p^{-1}(N_\beta)\to\R^*$$
 given by $m\mapsto p'(m)$.
Thus we obtain
 the matrix
\begin{equation}
\fP=
\begin{pmatrix}
\fp_{11}&\dots &\fp_{1t} &\fp_{1\infty}\\
\vdots&\ddots&\vdots&\vdots  \\
\fp_{s1}&\dots &\fp_{st} &\fp_{s\infty}\\
\fp_{\infty 1}&\dots &\fp_{\infty t} &\fp_{\infty\infty}
\end{pmatrix}
\end{equation}
consisting of measures on the group $\R^*$.
These measures satisfy the following equalities
\begin{equation}
\sum_{j=1}^t\int_{\R^*}d\fp_{ij}(x)
+\int_{\R^*}d\fp_{i\infty}(x)=\mu_i,
\qquad i=1,2,\dots, s
,\end{equation}
\begin{equation}
\sum_{i=1}^s\int_{\R^*}x\,d\fp_{ij}(x)+
\int_{\R^*}x\,d\fp_{\infty j}(x)=\nu_j,
\qquad j=1,2,\dots,t
,\end{equation}
and  the  conditions
\begin{equation}
\fp_{ij},\fp_{i\infty}, \fp_{\infty j}\in\cM^\tr,
\qquad
\fp_{\infty\infty}\in\cM^\chtr
.\end{equation}

Obviously, the map $p\mapsto\fP$
is constant on each double coset
$$
\Ams_\infty(M\bigl|\,\cU)\setminus\Gms_\infty/
\Ams_\infty(M\bigl|\,\cV)
$$
and moreover this defines a bijection
between the double coset space
and the space of all the matrices (11)
satisfying (12)--(14).

We also will write the matrix (11) in the
$(s+1)\times (t+1)$-block form
$$
\fP=\begin{pmatrix}
\fP_{\fin,\fin}&\fP_{\fin,\infty}\\
\fP_{\infty,\fin}&\fp_{\infty,\infty}
\end{pmatrix}
.$$

\smallskip

{\bf 2.6. Product of double cosets.}
Now consider 3 partions of the space $M$
(all these partitions can coincide)
\begin{gather*}
\cU: M=M_1\cup\dots\cup  M_s\cup M_\infty,\qquad
  \cV: M=N_1\cup\dots\cup
  N_t\cup N_\infty,\\
\cW: K=K_1\cup\dots\cup  K_r\cup K_\infty
.\end{gather*}
We intend to define the multiplication
of the double cosets
\begin{multline*}\!\!\!\!\!\!
\Ams_\infty(M\bigl|\,\cU)\setminus\Gms_\infty/
\Ams_\infty(M\bigl|\,\cV)
\,\, \times \,\,
\Ams_\infty(M\bigl|\,\cV)\setminus\Gms_\infty/
\Ams_\infty(M\bigl|\,\cW)\to\\
\to
\Ams_\infty(M\bigl|\,\cU)\setminus\Gms_\infty/
\Ams_\infty(M\bigl|\,\cW),
\end{multline*}
i.e., we want to define a multiplication
of matrices (11).

We say that a sequence
\begin{multline*}
h_n=(h_n^{(1)},\dots,h_n^{(t)}, h_n^{(\infty)})\in
\\ \in
\Ams_\infty(M\bigl|\cV)=
\Ams(N_1)\times\dots\times\Ams(N_t)\times
\Ams_\infty(N_\infty)
\end{multline*}
is {\it generic}
if all the sequences $h_n^{(\beta)}$
are generic ($\beta=1,2,\dots,t,\infty$).

Now we repeat the double coset multiplication
construction.
Consider a matrix
$\fP$, which corresponds to some
element
of $\Ams_\infty(M\bigl|\,\cU)\setminus\Gms_\infty/
\Ams_\infty(M\bigl|\,\cV)$.
Consider
 a matrix $\fQ$, which corresponds to some element
of
$\Ams_\infty(M\bigl|\,\cV)\setminus\Gms_\infty(M)/
\Ams_\infty(M\bigl|\,\cW)
$.
Consider the representatives $p,q\in\Gms_\infty(M)$
of these double cosets.
For a generic sequence
$h_n\in\Ams_\infty(M\bigl|\,\cU)$,
denote by $\fR_n$ the
element of the double coset
space
$$
\Ams_\infty(M\bigl|\,\cU)\setminus\Gms_\infty(M)/
\Ams_\infty(M\bigl|\,\cW),
$$
containing $qh_np$. Then the limit
$\fR$ of $\fR_n$ is given by
\begin{equation}
\fR=\begin{pmatrix}
\fQ_{\fin,\fin}\cdot D\cdot \fP_{\fin,\fin} &
\fQ_{\fin,\fin}\cdot D\cdot \fP_{\fin,\infty}+
    \fQ_{\fin,\infty}\\
\fQ_{\infty,\fin}\cdot D\cdot \fP_{\fin,\fin}
+\fP_{\infty,\fin}&
\fQ_{\infty,\fin}\cdot D\cdot \fP_{\fin,\infty}+
  \fP_{\infty,\infty}+\fQ_{\infty,\infty}
  \end{pmatrix}
,  \end{equation}
where
$$
D=\begin{pmatrix}
\nu_1^{-1}& 0& \dots\\
0& \nu_2^{-1} &\dots\\
\vdots & \vdots &\ddots
\end{pmatrix}
,$$
and $\nu_j$ are the measures
of  the elements $N_1$,\dots, $N_t$
of the partition $\cV$.

The associativity of the product can be
easily checked by
a direct calculation.

{\bf 2.7. $\star$-Polymorphisms of finite
bordered spaces.}
Let $M^\star$, $N^\star$ be {\it finite}
bordered spaces,
 let the measures of (finite) points
be $\mu_1$, $\mu_2$, \dots, $\mu_s$
and $\nu_1$, $\nu_2$, \dots, $\nu_t$.
An  element $\fP$ of $\Poll^\star(M^\star,N^\star)$
(a $\star$-polymorphism)
is a $(s+1)\times (t+1)$-matrix
(11) satisfying the conditions (12)--(14).
The product of the polymorphisms is given by
(15).

\smallskip

{\bf 2.8. $\star$-Polymorphisms
of general bordered spaces.}
Let $(M^\star,\mu)$, $(N^\star,\nu)$ be
bordered spaces.
An element $\fP$ of
$\Poll^\star(M^\star,N^\star)$
is a
measure on $M^\star\times N^\star\times \R^*$
satisfying  some conditions given below.
It is natural to represent
this measure as the block matrix
$$
\fP=\begin{pmatrix}
\fP_{\fin,\fin}&\fP_{\fin,\infty}\\
\fP_{\infty,\fin}&\fP_{\infty,\infty}
\end{pmatrix}
,$$
where
$\fP_{\fin,\fin}$ is a measure on
$M\times N\times\R^*$,
$\fP_{\fin,\infty}$ is a measure on
$\xi^M_\infty \times N\times\R^*\simeq N\times\R^*$,
$\fP_{\infty,\fin}$ is a measure on
 $M\times\xi^N_\infty \times \R^*
 \simeq M \times \R^*$,
and $\fP_{\infty,\infty}$ is a measure on
 $\xi^M_\infty \times\xi^N_\infty \times \R^*
 \simeq \R^*$.

The measure $\fP$ satisfies the following
conditions (which repeat the conditions (12)--(14)).

1. $\fP_{\fin,\fin}$,
$\fP_{\fin,\infty}$,
$\fP_{\infty,\fin}\in \cM^\tr$, and
$\fP_{\infty,\infty}\in\cM^\chtr$.

2. Let us restrict the measure $\fP$
to the set $M \times N^\star \times \R^*$.
Then the image of  this restriction
under the map
$M \times N^\star \times \R^*\to M$
coincides with $\mu$.

3. Let us restrict the measure $x\cdot\fP$
to the set $M^\star \times N\times \R^*$.
Then the image of this restriction
under the map
$M^\star \times N\times \R^*\to N$
coincides with $\nu$.

\smallskip

Product of $\star$-polymorphisms
is defined by the same formula (15).
We only must define the products in each block.
For instance,
let us give
an interpretation
of $\fQ_{\fin,\fin}\cdot D\cdot \fP_{\fin,\fin}$.
It is sufficient to use the prescription
1.8. To avoid a divergence,
consider countable partions of $M,N,K$ into
pieces with finite measures.
Then we consider approximative sequences
of partitions etc.etc.

\smallskip

{\bf 2.9. Embedding $\Gms_\infty\to \Poll^\star$.}
For $q\in\Gms_\infty(M)$, we define
the matrix $\fQ$ by the conditions
$$\fQ_{\fin,\infty}=0,\qquad\fQ_{\infty,\fin}=0,
\qquad
\fQ_{\infty,\infty}=0,$$
and $\fQ_{\fin,\fin}$ is the image of
the measure
$\mu$ under the map $M\to M\times M\times \R^*$
given by $m\mapsto (m,q(m), q'(m)).$

\smallskip

{\sc Remark.} The group $\Gms_\infty(M)$
is exactly the group of all invertible
$\star$-polymorphisms
of $M^\star$.


Below {\it we identify elements of $\Gms_\infty(M)$
and the corresponding elements of
$\Poll^\star(M^\star,M^\star)$.
}

\smallskip

{\bf 2.10. A remark on the formula
for product.}
 The exotic multiplication (15) of matrices
is a degeneration of the usual
matrix multiplication.
Indeed, let $\epsilon$ be infinitely small.
Consider block $(n+1)\times(n+1)$-matrices
 having the form
$$
\begin{pmatrix}
A+o(1)& \epsilon b+o(\epsilon)\\
\epsilon c+o(\epsilon)& 1+\epsilon^2 d+
o(\epsilon^2)
\end{pmatrix}
.$$
Then the product of such matrices
has the form
\begin{multline}
\begin{pmatrix}
A+o(1)& \epsilon b+o(\epsilon)\\
\epsilon c+o(\epsilon)& 1+\epsilon^2 d+
o(\epsilon^2)
\end{pmatrix}
\begin{pmatrix}
A'+o(1)& \epsilon b'+o(\epsilon)\\
\epsilon c'+o(\epsilon)& 1+\epsilon^2 d'+
o(\epsilon^2)
\end{pmatrix}=\\=
\begin{pmatrix}
AA'+o(1)& \epsilon (Ab'+b)+o(\epsilon)\\
\epsilon (cA'+c')+o(\epsilon)&
1+\epsilon^2 (d+d'+cb')
+o(\epsilon^2)
\end{pmatrix}
,\end{multline}
and we obtain the formula similar to (15).

\smallskip

{\bf 2.11. Remarks on
the convergence in $\Poll^\star$.}
Let $\fP\in\Poll^\star(M^\star$, $N^\star)$.
Let $A\subset M^\star$, $B \subset N^\star$
be measurable subsets (see 2.1).
We restrict the measure $\fP$
to $A\times B\times \R^*$.
Denote by $\fp[A\times B]$ the image
of this restriction under the projection
$A\times B\times \R^*\to \R^*$.

Let
$\fP_j, \fP\in\Poll^\star(M^\star$, $N^\star)$.
The sequence $\fP_j$ converges to $ \fP$
if the following two conditions
are satisfied.

a) For any measurable subsets
$A\subset M^\star$, $B \subset N^\star$
the sequence $\fp_j[A\times B]$
converges to $\fp[A\times B]$
in $\cM^\chtr$.

b) Moreover, if $A$ or $B$ does not contain
$\xi_\infty$, then we have convergence in
$\cM^\tr$.

\smallskip

{\sc Examples.}
a) Let $M=\R$. Let $y=f(x)$ be a diffeomorphism
of $\R$.
Assume
$$q_n(x)=f(x)+n.$$
Let us describe the limit $\fP$ of the sequence
$q_n$ in the sense of
$\Poll^\star(\R^\star,\R^\star)$.
The measure $\fP$ is supported by
$\R\times\xi_\infty\times\R^*$.
It coincides with the image of the Lebesgue
measure on $\R$ under the map
$$x\mapsto (x,\xi_\infty, f'(x)).$$

b) Under the same conditions,
the limit $\fQ$
of the sequence
$$q_n(x)=f(x-n)+n$$
is supported by
$\xi_\infty \times\xi_\infty\times\R^*$.
It coincides with the image of the Lebesgue
measure under the map
$$x\mapsto (\xi_\infty,\xi_\infty, f'(x)).$$

It is easy to
understand that
{\it the group $\Gms_\infty$
is dense in the semigroup
$\Poll^\star(M^\star,M^\star)$.}
Thus this semigroup
is some kind of a boundary
of the group $\Gms_\infty(M)$.

\smallskip

{\bf 2.12. Remarks on the polymorphisms
related to $\Ams_\infty$.}
We discuss this case for completeness.
Let $A^\star$, $B^\star$ be bordered spaces.
A polymorphism $\fP$ is a measure on $A\times B$
satisfying the conditions

1. The projection of $\fP$ onto $A$ is
majorized by $\alpha$

2. The projection of $\fP$ onto $B$
is majorized by $\beta$.

We define the product of polymorphisms by
the same formula (1).

\smallskip

{\sc Remark.}
We also can define this type of polymorphisms
$A^\star\to B^\star$ as $\star$-polymorphisms
supported by the set
$$
A^\star\times B^\star\times 1\subset
A^\star\times B^\star\times \R^*
$$

\medskip

{\large\bf 3. Construction
of functor}

\medskip

{\bf 3.1. Configurations.}
We say that a {\it configuration}
on a bordered space $M^\star$
is a countable (or finite) collection
\begin{equation}
\bm=\left[\begin{matrix}
m_1,&m_2,&m_3,\dots\\
p_1,&p_2,&p_3,\dots
\end{matrix}\right]
\end{equation}
 of distinct points $(m_1,m_2,m_3,\dots)$
 of $M^\star$
having integer positive
 multiplicities $p_1,p_2,p_3,\dots$.
We also assume that any configuration
contains $\xi_\infty$
with infinite multiplicity.
The collection $\bm$ is defined up to permutations
of the points together with their multiplicities
(i.e., up to the permutations of columns of (17)).

We also will give another definition.
A {\it configuration} is a map
$\phi$ from  a countable
set
$Z$ to $M^\star$ such that the preimage
$\phi^{-1}(\xi_\infty)$
of $\xi_\infty$ contains infinite number of points.
Two configurations $\phi:Z\to M^\star$,
$\phi':Z'\to M^\star$ are {\it equivalent} if
there exists a bijection $\psi:Z\to Z'$
such that $\phi=\phi'\psi$.

Of course, these two definitions are equivalent.
Indeed, consider a map $\phi:Z\to M^\star$.
The set $(m_1,m_2,\dots)$
is the image of $\phi$;
the multiplicity $p_j$ of a point
$m_j$ is number of elements in
$\phi^{-1}(m_j)$.

\pic

\smallskip

Denote by $\Omega(M^\star)$
 the space of configurations
 on $M^\star$ defined up to equivalence.

For a map $\rho:M^\star\to N^\star$
(we assume $\rho(\xi_\infty)=\xi_\infty$),
we have the natural map
$\Omega(M^\star)\to \Omega(N^\star)$
given by
$\phi\mapsto \rho\circ\phi$.
In particular, for any partition $\cU$,
we obtain the map
$\Omega(M)\to\Omega(M/\cU)$.

\smallskip

{\bf 3.2. Poisson measures: finite case.}
Consider a finite
bordered  space $M^\star$,
let the measures of the points
$m_1$, \dots, $m_k$ of $M$ be
$\mu_1$, \dots, $\mu_k$.
For
 a configuration $\phi: Z\to M^\star$,
we denote by $p_j$
the number of points in the preimage
 $\phi^{-1}(m_j)$ (the {\it multiplicity}
 of $m_j$, it can be 0).
Thus the space $\Omega(M^\star)$ is identified
with the space $\Z_+^k$.
We define the Poisson measure $\nu_M$
on $\Omega(M^\star)$ by the condition:
  the
measure of the point $(p_1,p_2,\dots,p_k)\in\Z_+$
is
$$\prod_k \frac{\mu_k^{p_k}e^{-\mu_k}}{p_k!}.$$

\smallskip

{\bf 3.3. Poisson measures: general case.}
The Poisson measure $\nu_M$
on $\Omega(M)$ is defined by the condition:
for any partition $\cU$ of $M^\star$,
the image of $\nu_M$ under the map
$$\Omega(M^\star)\to\Omega(M^\star/\cU)$$
coinsides with the
Poisson measure $\nu_{M^\star/\cU}$
on
$\Omega(M^\star/\cU)$,
see
\cite{Kin}, \cite{VGG}, \cite{Nerb}, \cite{Nerp}
for more details.

\smallskip

{\bf 3.4. Normed exponent.}
Consider a measure $\psi\in\cM^\chtr$ on
$\R^*$.
The function
$$r(s)=\int_{\R^*} (x^{is}-1)\,d\psi(x)$$
is a
well-defined conditionally
positive definite function on $\R$. Hence
$e^{r(s)}$
is a positive definite function.
Hence $e^{r(s)}$ is a Fourier transform
of some measure $\kappa$
$$e^{r(s)}=\int_0^\infty x^{is} d\kappa(x).$$
 We define the {\it normed exponent} $\exp_\circ[\psi]$
 by
 $$\exp_\circ[\psi]=\kappa.$$

 {\sc Remark.}
 Assume $\psi\in\cM^\tr$. Then
$$\exp_\circ[\psi]=\exp\Bigl\{-\int_{\R^*}d\psi(x)\Bigr\}
\cdot\Bigl\{\delta_1+\frac{\psi}{1!}+
\frac{\psi*\psi}{2!}+
\frac{\psi*\psi*\psi}{3!}+\dots
\Bigr\},
$$
where $\delta_1$ denotes the atomic unit
measure
supported
by the point $1\in\R^*$.

\smallskip

{\bf 3.5. Partial bijections.}
Let $S,T$ be finite sets. A partial
bijection $Q:S\to T$ is a bijection
of a subset $A\subset S$ to a subset $B\subset T$.
We say that $A$ is the {\it domain}
of $Q$ (the notation is $A=\Dom(Q)$)
and $B$ is the {\it image} of $Q$
(the notation is
$B=\Im(Q)$). We denote by $\PB(S,T)$
the set of
all partial bijections $S\to T$.

\smallskip

{\bf 3.6. The construction.}
Consider  finite spaces $M^\star$, $N^\star$
and the associated spaces
$\Omega(M^\star)$, $\Omega(N^\star)$ equipped with
the Poisson measures.

 For each $\star$-polymorphism
 $\fP\in\Poll^\star(M^\star,N^\star)$,
 we will construct an $\R^*$-polymorphism
 $\omega(\fP)\in\Pol(\Omega(M^\star),\Omega(N^\star))$.

 \smallskip

 Consider arbitrary configurations
 $\phi: Z\to M^\star$ and $\psi:Y\to N^\star$.
 Denote by $S\subset Z$, $T\subset Y$ the preimages
 of the sets $M$, $N$; obviously, the configuration
 $\phi$ (resp. $\psi$)
 is completely defined by the restriction to S
 (resp. $T$).
Denote by $p_i$ (resp $q_j$)
 the multiplicities
of the points of the configuration $\phi$
(resp. $\psi$).
We define the measure $\omega_{\phi\psi}$
on $\R^*$ by
\begin{multline}
\omega_{\phi\psi}=
C\cdot\delta_h * \exp_{\circ}
[\fp_{\infty\infty}]\,*
\frac{1}{\prod p_i! \,\prod q_j!}
\sum\limits_{Q\in \PB(S,T)}
\Bigl\{
\prod\limits_{s\in\Dom (Q), t=Qs}
\fp_{st}\, *\,\\ *
\prod\limits_{s\not\in\Dom(Q)} \fp_{s\infty}
\,*\,\prod\limits_{t\not\in \Im(Q)}
    \fp_{\infty t}\Bigr\}
,    \end{multline}
where the summation is given
over the set of all
partial
bijections $T\to S$,

\picc

\noindent $\exp_\circ[\cdot]$ denotes the normed
    exponent,
symbols $\prod$ denote convolutions
of measures on $\R^*$,
the scalar factor $C$ is given by
$$
C=\exp\Bigl\{-\sum\limits_{i,j}
\int d\fp_{ij}-\sum_i \int d\fp_{i\infty}
-\sum_j \int d\fp_{\infty j}\Bigr\}
,$$
and $\delta_h$ is the unit $\delta$-measure
on $\R^*$ supported by
the point
$$
h=\exp\Bigl\{-
\int(x-1)\,d\bigl[\sum\limits_{i,j}\fp_{ij}+
\sum_i  \fp_{i\infty}+\sum_j \fp_{\infty j}+
\fp_{\infty\infty}\bigr]\Bigr\}
.$$

{\sc Theorem A.}
a) {\it The matrix
$\omega(\fP)$ composed from  measures $\omega_{\phi\psi}$
is an element of $\Pol(\Omega(M^\star),\Omega(N^\star))$.
}

b) {\it The map $\fP\mapsto\omega(\fP)$
is a functor, i.e.
for each finite
bordered spaces $M^\star$, $N^\star$, $K^\star$
and for each
$\star$-polymorphisms
$\fP\in\Poll^\star(M^\star, N^\star)$,
$\fQ\in\Poll^\star(N^\star,K^\star)$,}
\begin{equation}
\omega(\fQ)\omega(\fP)=\omega(\fQ\fP)
.\end{equation}

{\bf 3.7. Construction of the functor $(\Omega,\omega)$
in general case.}
Let $(M^\star,\mu)$, $(N^\star,\nu)$ be arbitrary
bordered spaces, and let
$$\cU:M^\star=M_1\cup\dots\cup M_k\cup  M_\infty,
\qquad \cV:N^\star=N_1\cup\dots\cup N_l\cup N_\infty$$
be  partitions
of $M^\star$, $N^\star$ respectively.
Let $\fP\in\Poll^\star(M^\star,N^\star)$.
For any $\alpha=1,\dots,k,\infty$
and $\beta=1,\dots, l,\infty$
we consider the map
$$M_\alpha\times N_\beta\times \R^*\to \R^*$$
and the image of the measure $\fP$ under this map.
Thus we obtain an
$\cM$-valued matrix, it defines the
element of
$\Poll^\star(M^\star/\cU,N^\star/\cV)$.
We denote it by
$$\fP^\downarrow_{[\cU,\cV]}.$$

Let $M^\star$ be a bordered space.
Let $\cU^{(j)}$ be a sequence of partitions,
and let each $\cU^{(j+1)}$ be a refinement
of $\cU^{(j)}$.
We say that the sequence $\cU^{(j)}$
of partitions
is {\it approximative} if finite elements of
all partitions $\cU^{(j)}$
generate the Borel $\sigma$-algebra of $M$.

Fix
$\fP\in\Poll^\star(M^\star,N^\star)$.
 Let $\cU^{(j)}$, $\cV^{(j)}$
be approximative sequences of partitions
of $M^\star$, $N^\star$ respectively.
Then we have
the chain of the
spaces
\begin{equation}
\dots\leftarrow M^\star/\cU^{(j)}
\times N^\star/\cV^{(j)}\times\R^*
\leftarrow M^\star/\cU^{(j+1)}
\times N^\star/\cV^{(j+1)}\times\R^*
\leftarrow \dots
 \end{equation}
The sequence
$\fP^\downarrow_{[\cU^{(j)},\cV^{(j)}]}$
of bordered polymorphisms
(defined in 3.1)
is a projective sequence of measures
with respect to the chain (20).

\smallskip

 {\sc Theorem B.}
a)
{\it Let $\fP\in\Poll^\star(M^\star,N^\star)$.
Let
 $\cU^{(j)}$, $\cV^{(j)}$
be approximative sequences of partitions
of $M^\star$, $N^\star$.
Then the system
$$
\omega(\fP^\downarrow_{[\cU^{(j)},\cV^{(j)}]})
\in\Pol
\bigl(\Omega(M^\star/ \cU^{(j)}),
\Omega(M^\star/ \cV^{(j)})\bigr)
$$
 is a projective
 system of measures with respect to the maps
$$\dots\leftarrow
\Omega(M^\star/ \cU^{(j)})
\times \Omega(N^\star/ \cV^{(j)})
\leftarrow
 \Omega(M^\star/ \cU^{(j+1)})
\times \Omega(N^\star/ \cV^{(j+1)}) \leftarrow\dots $$
The inverse limit
$$\omega(\fP)\in\Pol(\Omega(M),\Omega(N))$$
of this chain
does not depend on the choice
of the approximative sequences
$\cU^{(j)}$ and $\cV^{(j)}$.}

\smallskip

  b) {\it The map $\fP\mapsto\omega(\fP)$
  is a functor, i.e.,
  for each $M^\star$, $M^\star$, $K^\star$
and each $\star$-polymorphisms
 $\fP\in\Poll^\star(M^\star, N^\star)$,
$\fQ\in\Pol(N^\star,K^\star)$,}
$$\omega(\fQ)\omega(\fP)=\omega(\fQ\fP).$$

\smallskip

c) {\it Let $q\in\Gms_\infty(M)$. Then
$\omega(q)$ is the transformation of
$\Omega(M^\star)$ given by
$(m_1,m_2,\dots)\mapsto (q (m_1),q(m_2),\dots)$,
i.e., our functor $(\Omega,\omega)$ extends the map}
(4).

\smallskip

{\bf 3.9. Remarks on the proofs.}
There are two ways to prove Theorem A.
The both  ways require
some calculations.

The first way. Consider
 a space $M^\star$ with a continuous infinite
 measure. Consider a partition $\cU$
 of $M^\star$.
 We have a map from $\Omega(M^\star)$
 to the countable space $\Omega(M^\star/\cU)$,
 and thus we have a partition of
 the space $\Omega(M^\star)$.
 Denote this partition by $\Omega(\cU)$.
 For any partition $\cU$, the map
 $$\Gms_\infty(M)\to\Gms(\Omega(M))$$
 induces the maps of the subgroups
\begin{equation}
 \Ams_\infty(M\bigl|\,\cU)\to
     \Ams\bigl(\Omega(M^\star\bigl| \Omega(\cU)\bigr)
.\end{equation}
 Thus we have the map of double cosets
\begin{multline*}
\Ams_\infty(M\bigl|\,\cU)\setminus\Gms_\infty(M)/
\Ams_\infty(M\bigl|\,\cV)\to
\\
\to
     \Ams\bigl(\Omega(M^\star\bigl| \Omega(\cU)\bigr)
     \setminus
\Gms(\Omega(M))/
     \Ams\bigl(\Omega(M^\star\bigl| \Omega(\cV)\bigr)
.\end{multline*}

The map (21) transforms generic sequences to
generic sequences,
and this implies the product formula (19).
For obtaining (18),
it remains to calculate this map explicitly.

Another way of proof of (19) is a direct calculation.
The formula (19) is equivalent
to a family of identities
for some infinite sums depending
on elements of $\cM^\tr$.
The same identities for series
depending on complex variables
appear in the following situation.

Consider the space ${\cal F}_n$ of entire
functions on $\C^n$.
Let $A:\C^n\to\C^n$ be a linear operator,
let $b,c \in \C^n$. Consider
the linear operator
$$U(A,b,c)f(z)=f(Az+b)\exp(\sum c_j z_j).$$
Obviously,
\begin{equation}
U(A,b,c)U(A',b',c')=
\exp(\sum b_j c_j')
U(A'A, A'b+b', A^tc'+c)
.\end{equation}
Consider the matrix elements
of this operator in the basis
$z_1^{p_1}\dots z_n^{p_n}$.
The explicit expresions for
these matrix elements can be easily written
as polynomial on $A,b,c$;
they almost coincide with the expresion (18).
In this basis,
the product formula (22)
 is some collection of identities
for  series of complex numbers.

The identities that are necessary for
the proof
of (19) are the same, but the complex numbers
are replaced by elements of the semigroup
$\cM^\tr$.
It remains to observe, that
for any $s\in\C$, such that
$0\le\Re\, s\le 1$,
the map
$$
\fu\mapsto \int_0^\infty
   x^s d\fu(x)
   $$
   is a homomorphism of rings $\cM^\tr\to \C$
   and this family of homomorphisms
   separates
  elements of $\cM^\tr$.

Theorem B is a corollary of
Approximation Theorem
for categories  \cite{Nerb}, Theorem 8.1.10.

\sf

Address (autumn 2001):

     Erwin Schr\"odinger Institute for Mathematical
             Physics,

    Boltzmanngasse, 9, Wien 1020, Austria

    \smallskip

    Permanent address:

   Math.Physics Group,

   Institute of Theoretical and Experimental Physics,

     Bolshaya Cheremushkinskaya, 25,

     Moscow 117259

     Russia

     \smallskip

    {\tt e-mail neretin@main.mccme.rssi.ru}

  \end{document}